\documentclass[aps,prb,reprint,groupedaddress]{revtex4-1}
\usepackage{graphics}

\begin{document}

%\preprint{}

\title{The 2D percolation transition in two atomic layers of Fe on W(110): \\Direct measurement of a static percolation critical exponent in a 2D Ising system}


\author{K. Dixon}
%\thanks{}
\author{D. Venus}
\email{[corresponding author] venus@physics.mcmaster.ca}
\affiliation{Department of Physics and Astronomy, McMaster University, Hamilton, Ontario, Canada}

\date{\today}

\begin{abstract}
When the coverage of the second atomic layer of Fe in an Fe/W(110) ultrathin film reaches a critical value, the system moves suddenly from a frustrated magnetic state without long-range order to an in-plane ferromagnetic state with long-range order, and displays many features of a percolation transition.  Measurements of the magnetic susceptibility as the films are grown at 255 K show power law scaling that is limited by noise at low deposition, and by the dynamics of the paramagnetic, frustrated state at high deposition.  Because the measurements represent a system driven by a finite field oscillating at a finite frequency, it is demonstated that the threshold deposition for percolation is bounded by the depositions where the real and imaginary components of the susceptibility have maxima.  Fitting for the critical exponent of the static susceptibility at these bounds gives a bounded value for $\gamma_p=2.39\pm0.04$, in agreement with theory.
\end{abstract}
%\keywords{}

\maketitle

\section{Introduction}
Despite a long history, the percolation phase transition continues to be an active and relevant field of research.   As materials physics moves increasingly to nanoscale systems and ultrathin film structures, the importance of the long-range connectivity of the structures, and how this affects their material properties, has become more prominent.   Examples include the growth of an ultrathin film\cite{Chinta,Yu} and its material and transport properties\cite{Wagner,Tsunado}.  In addition, recent theoretical interest has been motivated by the finding that percolation (or dilution) can create unusual behaviour in the quantum phase transition of the 2D Ising model\cite{Senthil}.   Although the most striking results are restricted to $T=0$, or exponentially close to zero\cite{Vojta}, the treatment of the percolation transition at zero temperature as a multicritical point implies that quantum fluctuations may influence the nature of the phase transition at finite temperature.\cite{Vojta2}  Thus there is a search for experimental systems in which these effects might be exhibited.\cite{Vajk}

Given the practical importance of percolation, and the efforts to detect departures from the classical values of the percolation exponents due to quantum effects, it would be natural to expect that robust experimental measurements have confirmed the classically predicted behaviour at finite temperature.  However, this seems not to be the case, even for the ``textbook" case of 2D percolation of a 2D Ising model system.\cite{StaufferAharony}   The foundational experimental study of this universality class used neutron scattering\cite{Cowley,Birgeneau2} to investigate the quasi-2D diluted Ising antiferromagnet $\mathrm{Rb_2Co_cMg_{1-c}F_4}$.  These experiments employed a series of 3D samples of fixed dilution, $c$, close to the percolation threshold, and determined the critical exponent of the correlation length as a function of temperature, $\nu_T$, through the temperature variation of the magnetic scattering.  They then inferred that the crossover exponent $\phi=\nu_T/\nu_p=1$ by comparison to the theoretical value of $\nu_p$, the critical exponent of the correlation length as a function of dilution.  Through further analysis of the correlation length as a function of temperature in the paramagnetic phase, they investigated $\gamma_T$, the critical exponent for the susceptibility (mean island size in percolation) as a function of temperature.  They reported three disparate values for three samples with different dilutions, and declined to reach a conclusion. Rather, the authors noted that an ``optimistic extrapolation" to the expected percolation concentration yielded a value in agreement (within the uncertainty) with the theoretical value expected if $\phi=1$.

This important study was limited by the inevitable difficulty in studying percolation as a function of dilution in 3D samples of fixed composition.   In this regard, an ultrathin film is the ideal system in which to study 2D percolation, since the film can be continuously monitored and followed through the percolation transition as it is deposited.  Thus continuous measurements can be made parallel to both the temperature and the concentration axes.  Yet, to our knowledge, the only experimental determination of a static percolation exponent as a function of deposition for a 2D film is early work summarized in ref. \onlinecite{Vaz}.  These experiments on ultrathin Co/Cu\cite{Schumann,Hope,Kupper} and Fe/InAs\cite{Tselepi} used hysteresis loops to determine a handful of data points that were fit to power law scaling. 

This situation has been partially addressed in recent experiments\cite{Belanger} studying Fe/W(110) ultrathin films, which has been clearly shown to be a 2D Ising system.\cite{Back,Dunlavy,Dunlavy2}  Studies of the structural and magnetic properties of Fe/W(110) films provide evidence of two distinct percolation transitions as a function of coverage,\cite{Elmers} one when the first layer percolates, and another when the second layer percolates.  For the latter case, experiments show that islands in the second atomic layer of Fe/W(110) have perpendicular anisotropy\cite{Elmers2,Weber}, likely due to the large strain induced by pseudomorphic growth.  It appears that neighbouring islands are coupled antiferromagnetically through some interaction that is mediated by the first atomic layer of Fe and the W substrate.  The coexistence of these perpendicularly-magnetized islands with the continuous in-plane magnetized first atomic Fe layer creates a frustrated state that shows no long-range magnetic order.  At the critical coverage, long-range in-plane magnetic order suddenly returns, and the Curie temperature observed in remanence increases rapidly with deposition.

It has recently been demonstrated\cite{Belanger} that this second transition can be robustly detected in magnetic susceptibility measurements as a function of temperature \textit{and} as a function of deposition.  In both cases, the transition was marked by a strong, narrow peak consistent with a second order phase transition.  These peaks were used to map out the phase transition line in the $(p, T)$ plane.  The transition was confirmed to be a percolation transition by quantitative comparison to the theoretical expression\cite{StaufferAharony} for the percolation transition line of a 2D Ising system.
	
The present article follows up these investigations and reports further measurements of the magnetic susceptibility as a function of deposition as the second Fe layer of the film percolates.   Growing the films very slowly increases the signal-to-noise ratio so that the data allow a direct experimental determination of the critical exponents $\gamma_p=2.39\pm0.04$,  consistent with an experimental finding for the crossover exponent $\phi=1$.  Section II outlines the experimental methods by which the real and imaginary parts of the susceptibility are measured in the presence of a small driving field.  Section III reviews relevant theory, presents the experimental results and outlines the method of analysis. The final section summarizes the findings.

\section{Experimental methods}

The experimental methods and procedures have been described in detail in a previous publication that investigated the phase transition line for this percolation transition\cite{Belanger}.  The chief difference in the present experiments is that the films were grown at a much slower rate.

The experiments were performed in an ultrahigh vacuum environment, with the substrate W(110) crystal cleaned by oxygen treatments and flashing to white heat.  Cleanliness of the crystal surface was established using Auger electron spectroscopy (AES) and low energy electron diffraction (LEED).  The substrate could be heated by electron bombardment or radiation, and cooled by a copper braid running to a liquid nitrogen reservoir.  The films were deposited by thermal evaporation from the tip of a pure Fe wire, with a  beam of evaporant created by a pair of collimating apertures\cite{Jones}.  Since a certain proportion of the evaporated atoms are ionized, a current of order nA was produced on the final aperture.  By monitoring this current, the Fe flux could be kept constant.

Under normal conditions for film growth, the monitor current can be calibrated to provide a measure of the total deposition\cite{He} in nA min/ML.  This works well for monitor currents larger than about 0.25 nA that produce deposition rates of about 5-7 min/ML.  However, the present studies required slow film growths where the monitor current was significantly less than 0.1 nA.  While this small current was found to be stable, a leakage current comprised some portion of the total current, and the leakage current varied from one day to the next.  This made calibration of the total deposition unreliable.  As a result, the susceptibility measurements are presented as a function of deposition time, $t$, rather than deposition $\theta$. ($\theta$, in ML, is the total deposition required to form a film of as many atomic layers \textit{if} the film indeed grew as complete layers.) When necessary, the results of the previous study\cite{Belanger} are used to calibrate the deposition as 1.24$\pm$0.4 ML at the peak in the susceptibility corresponding to the percolation transition at 255 K.  Due to the constant rate of deposition (especially during the restricted time period corresponding to the critical range of the transition), the variables $p$ for fractional coverage, $\theta$ for deposition, and the deposition time $t$ are all linearly related.  Substituting one for another involves only a change in the prefactor to the critical power law scaling.

\begin{figure}
\scalebox{.5}{\includegraphics{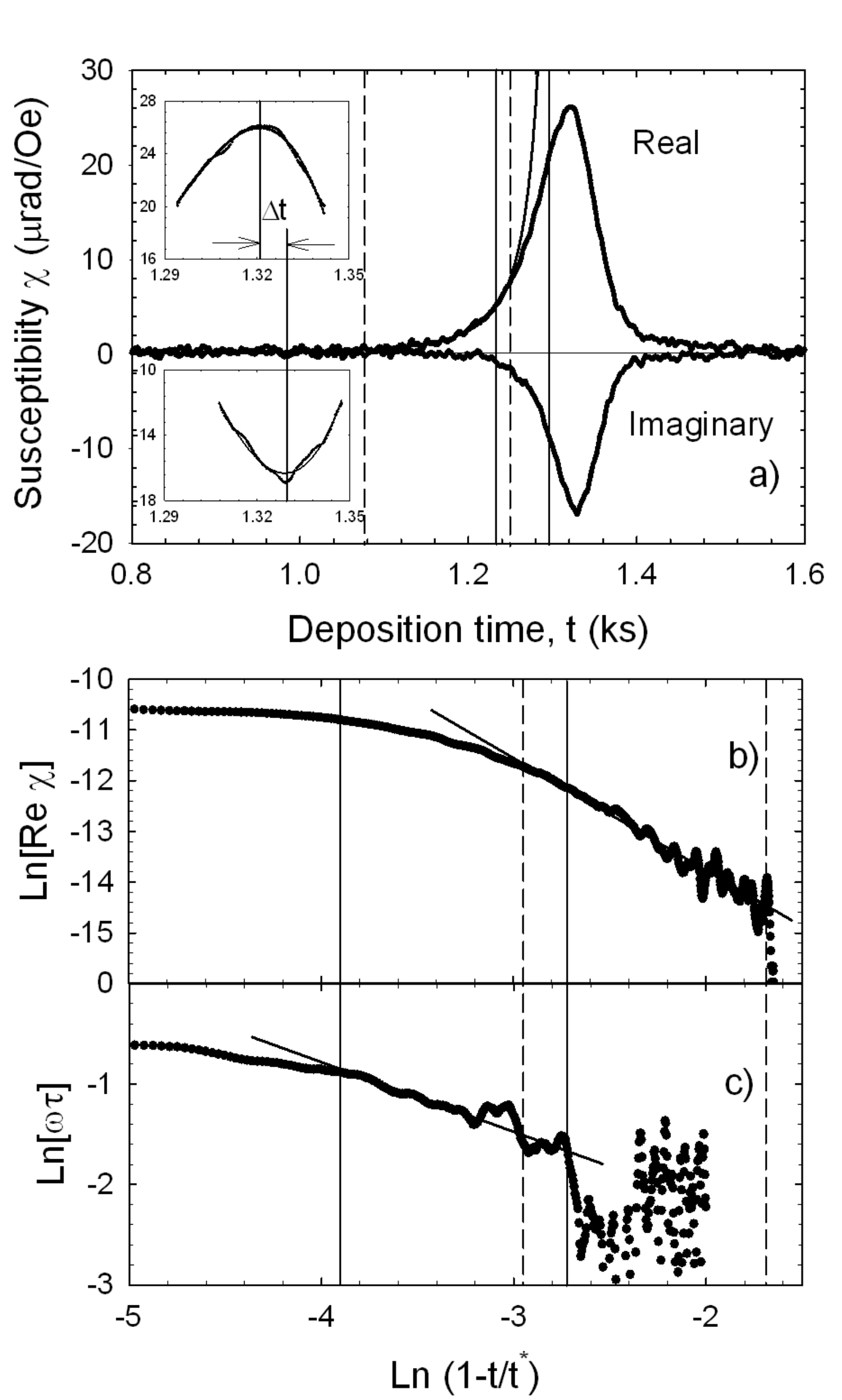}}
\caption{Magnetic susceptibility of a Fe/W(110) film measured as a function of time as the film was grown at a constant rate at 255 K.  a) The Real and Imaginary components are presented with opposite signs, consistent with eq.(\ref{dynamic}).  The insets have the same axis titles as the main figure, and show cubic fits to the Re and Im peaks to determine their locations and separation $\Delta t$.  b) Re$\chi$ is plotted on a ln-ln scale vs. reduced time, where $t^*$ is the deposition time corresponding to the coverage at percolation, $p^*$.  The dashed vertical lines (which appear in other parts of the figure) show the region of power law scaling.  The linear fit in the scaling region is shown by the solid line, and by the corresponding fitted line in part a).  c) The relaxation time from eq.(\ref{quotient}) is plotted on a ln-ln scale.  The apparent range of power law scaling is indicated by the solid vertical lines (which appear in other parts of the figure).  The dynamic scaling begins where the scaling of Re$\chi$ ends in part b), once $(\omega \tau)^2 \approx 0.06$.}
\label{sample_curve}
\end{figure}

The magnetic susceptibility was measured\cite{Arnold} \textit{via} the longitudinal magneto-optic Kerr effect (MOKE), using a HeNe laser beam reflected from the sample surface at close to 45$^o$.  The light passed through a polarizing crystal and a UHV window, reflected from the sample, passed through another UHV window and an analysing crystal that was almost crossed with respect to the polarizer, and struck a photodiode.  An optical compensation procedure ensured that the light exiting the second UHV window was linearly polarized\cite{Arnold2}.  This allowed a very small Kerr rotation of $<10$ nrad to be detected, and for the calibration of the susceptibility in absolute (non-magnetic) units of $\mu$rad/Oe.  The scattering plane of the light included the W($\overline{1}$10) surface direction, which is the easy axis for ultrathin Fe/W(110) films when they are magnetized in-plane\cite{Elmers6}.  A pair of air coils aligned with this direction produced a 210 Hz a.c. magnetic field at the crystal, and lock-in detection was used to isolate the portion of the photodiode signal at the frequency of the field.  This signal is proportional to the susceptibility;  the in-phase component corresponds to Re$\chi$ and the out-of-phase component to Im$\chi$.

In order to measure the susceptibility as the film grew, it was necessary to align the evaporator with the laser spot, which had a radius of about 0.7 mm.  This was accomplished by use of micrometers on a tripod attached to the evaporator.  AES measurements confirmed a uniform film thickness over a radius of at least 3 mm.  An example of a susceptibility measurement made during film growth, as the film passed through the percolation transition, is shown in fig.(\ref{sample_curve}a).  This measurement was made at a constant temperature of 255 K.


\section{Analysis and Experimental Results}
\subsection{Theoretical background}
The experimental determination of static critical exponents is particularly sensitive to the identification of the range over which critical scaling is exhibited, and of the transition point.  In fact, the largest source of uncertainty in the value of the exponent is usually the uncertainty in the value of the transition point.  For this reason, the following brief review pays particular attention to the effects of finite size, finite applied field, and measurement frequency effects that are present in all experiments.  

In the initial stages of growing an ultrathin film, deposited atoms fill a fraction $p$ of the substrate lattice sites and the atoms aggregate to form disconnected 2D islands.  With further growth, the islands begin to coalesce and at the critical fractional coverage $p_c$, corresponding to percolation, at least one island becomes effectively infinite in extent.  According to theory, the zero temperature percolation transition in a 2D system is a second order phase transition with universal critical exponents.\cite{StaufferAharony}  If the atoms have Ising magnetic moments interacting through a local exchange coupling, there is a mapping of a magnetic transition onto the percolation transition which preserves its universal properties.  It is assumed that the moments within the isolated islands are ferromagnetically aligned and that the islands themselves are too small to support internal magnetic domains -- that is, the system is superparamagnetic.  Then a paramagnetic-to-ferromagnetic transition accompanies percolation as the magnetic correlation length diverges.  As a result, the standard results of the critical scaling hypothesis for thermal transitions\cite{Stanley} can be applied directly by substituting the reduced coverage $\rho=(p_c-p)/p_c$ for the reduced temperature $\epsilon=(T-T_c)/T_c$.  For clarity, the critical exponents for the percolation transition are indicated by a subscript $p$ to distinguish them from the more familiar exponents for a thermal transition.  Note that this description applies directly to the paramagnetic phase, but that possible effects of magnetic domains must be considered in the  percolated, ferromagnetic phase.\cite{Aharoni} 

According to the critical scaling hypothesis,\cite{Stanley} the magnetic susceptibility, $\chi(\rho,H)$, measured in a magnetic field $H$, scales with coverage as a generalized homogenous function  $F$ of these two variables.  This can be expressed as a scaling with coverage
\begin{equation}
\chi(\rho,H)=\rho^{-\gamma_p} F[+1,\frac{H}{\rho^{\beta_p+\gamma_p}}],
\label{scale1}
\end{equation}
where $\gamma_p$ is the exponent of the susceptibility (corresponding to mean island size) and $\beta_p$ is the exponent of the magnetization (corresponding to the fraction of atoms in the infinite percolated island).  The index +1 in the first argument of $F$ correspond to the sign of $\rho$ in the paramagnetic state.  For an ideal system of infinite extent, measured in the limit $H\rightarrow0$, this  gives the familiar result
\begin{equation}
\chi(\rho)=F[+1,0]\rho^{-\gamma_p}\equiv\chi_+\rho^{-\gamma_p}.
\label{zeroT}
\end{equation}

Percolated systems have by their nature structural inhomogeneities, and the present measurements use a finite, low frequency field to drive the system.  In this case case, the magnetic susceptibility will not diverge, but will saturate through some combination of finite size and finite field effects.  If the percolation cascade effectively ends at a length scale $L$, then the susceptibility will saturate in the paramagnetic phase at $\rho_{max}=L^{-1/\nu_p}$. ($\nu_p$ is the critical exponent of the correlation length.)  It will still be possible to observe the approach to the characteristic divergence in eq.(\ref{scale1}), so long as $\rho > \rho_{max}$,  such that 
\begin{equation}
\frac{H}{\rho^{\beta_p+\gamma_p}}\rightarrow0^+.
\label{H_limit}
\end{equation}
An important question, that can only be answered through experiment, is what ranges of $H$ and $\rho$ will satisfy eq.({\ref{H_limit}}), but still remain in the critical region where eq.(\ref{zeroT}) applies.  

Even in a structurally ideal system, the susceptibility may saturate because of the finite size of the applied field.  The universal properties of the susceptibility can also be expressed as a scaling in the applied magnetic field:\cite{Stanley} 
\begin{equation}
\chi(\rho,H)=H^{-\gamma_p/(\gamma_p+\beta_p)} G[\frac{\rho}{H^{1/(\beta_p+\gamma_p)}},1],
\label{scale2}
\end{equation}
where $G$ is a different scaling function.  This relation is normally considered in the limit $\rho=0$, where it characterizes the critical ``isotherm" (iso-coverage (?) in the present case).  However, when a finite field is used to drive the system, the maximum of the susceptibility does not occur at $\rho=0$, but rather at the locus of points  $\rho_{max}(H)$ in the paramagnetic phase that meet the condition\cite{Gaunt,Kunkel}
\begin{equation}
\frac{\rho_{max}}{H^{1/(\beta_p+\gamma_p)}}=C,
\label{rhomax}
\end{equation}
where $C$ is a constant.  The saturated value of the susceptibility at these locations is
\begin{equation}
\chi(\rho_{max},H)=H^{-\gamma_p/(\gamma_p+\beta_p)}G[C,1].
\label{delta}
\end{equation}

The above analysis describes classical percolation at zero temperature.  Percolation at finite temperature is described using the theory of a bicritical point.\cite{StaufferAharony,Stanley2}   The percolation point at $(p_c, T=0)$ may be approached along the $T=0$ axis as a function of deposition, or along a continuous line of  phase transitions that connects it to the limit of a thermal transition of a complete 2D film at $(p=1, T=T_c)$.  The finite temperature phase transition line is given by $p^*(T)$, and is determined by the competition between the thermal correlation length $\xi_T$ and the percolation correlation length $\xi_p$, which diverge with critical exponents $\nu_T$ and $\nu_p$, respectively.  For a percolating 2D Ising magnetic system, the correlation lengths in the critical region are determined by fragile 1D chains of Ising spins that connect 2D islands to make larger islands.\cite{Coniglio}  A 1D Ising chain is a particularly straightforward geometry, as it requires only one site to break the chain.  Loosely speaking, the system is indifferent to whether the break is geometric (due to a missing atom) or magnetic (due to a thermally reversed Ising spin).  Because of this $\nu_T=\nu_p$ and the crossover exponent $\phi=1$.\cite{Coniglio}  This leads to the simple result that the finite temperature, equilibrium susceptibility $\chi_{eq}(\rho^*_T,H)$ is given by the zero temperature results outlined above, with $p_c$ replace by $p^*(T)$ so that
\begin{equation}
\rho \rightarrow \rho^*_T=\frac{p^*(T)-p}{p^*(T)}.
\label{rho*}
\end{equation}
In particular, the value of the critical exponent $\gamma_p$ does not depend upon the temperature of a path at constant temperature, so long as $p^*(T)-p_c$ is not ``too large".

Finally, it is necessary to account for the fact that the finite field used in the experiments oscillates at a finite frequency.  Dynamic scaling theory\cite{Hohenberg} makes the independent hypothesis that critical slowing down within the  paramagnetic phase follows universal behaviour, according to the dynamical model that is applicable.   Dynamic scaling proposes a characteristic time of an Ising system, $\tau$, that diverges as
\begin{equation}
\tau(\rho^*_T)=\tau_+ \; {\rho^*_T}^{-z_p \nu_p},
\label{zednu}
\end{equation}
where $\tau_+$ is the amplitude of the diverging relaxation time in the paramagnetic state, and $z_p$ is the dynamical exponent.  The explicit dependence of $\tau$ on $H$ has been suppressed under the assumption that if a deposition range can be found that satisfies the requirements of eq.(\ref{H_limit}) for $\chi_{eq}(\rho^*_T,H)$, it will also satisfy a similar requirement for $\tau$.  A number of theoretical\cite{Harris} and computational\cite{Biswal,Rammal} studies suggest that, if the magnetic dynamics is based upon the local, independent reversal of spins, the dynamical exponent of a 2D percolating Ising magnetic system at finite temperature is not universal, and that an effective value for $z_p$ is measured.  Whether or not this is the case, the ability of the system to relax to equilibrium depends upon the relative size of $\tau$ and the angular frequency $\omega$ of the measurement.  In the relaxation approximation, the susceptibility becomes\cite{Ogielski}
\begin{equation}
\chi(\rho^*_T,H)=\frac{1-i\omega \tau}{1+(\omega \tau)^2}\: \chi_{eq}(\rho^*_T,H).
\label{dynamic}
\end{equation}
This equation illustrates that driving the system creates a phase lag and results in an imaginary component of the susceptibility.  The real component of the susceptibility corresponds to the equilibrium susceptibility only when $(\omega \tau)^2 \ll 1$.

So long as the relaxation approximation applies, the dominant relaxation time of the system can be accessed by susceptibility measurements through the phase lag $\alpha$, where eq.(\ref{dynamic}) yields
\begin{equation}
\tan \alpha \equiv \frac{\mathrm{Im} \chi(\rho^*_T)}{\mathrm{Re} \chi(\rho^*_T)}= \omega \tau(\rho^*_T).
\label{quotient}
\end{equation}

\begin{figure}
\scalebox{.45}{\includegraphics{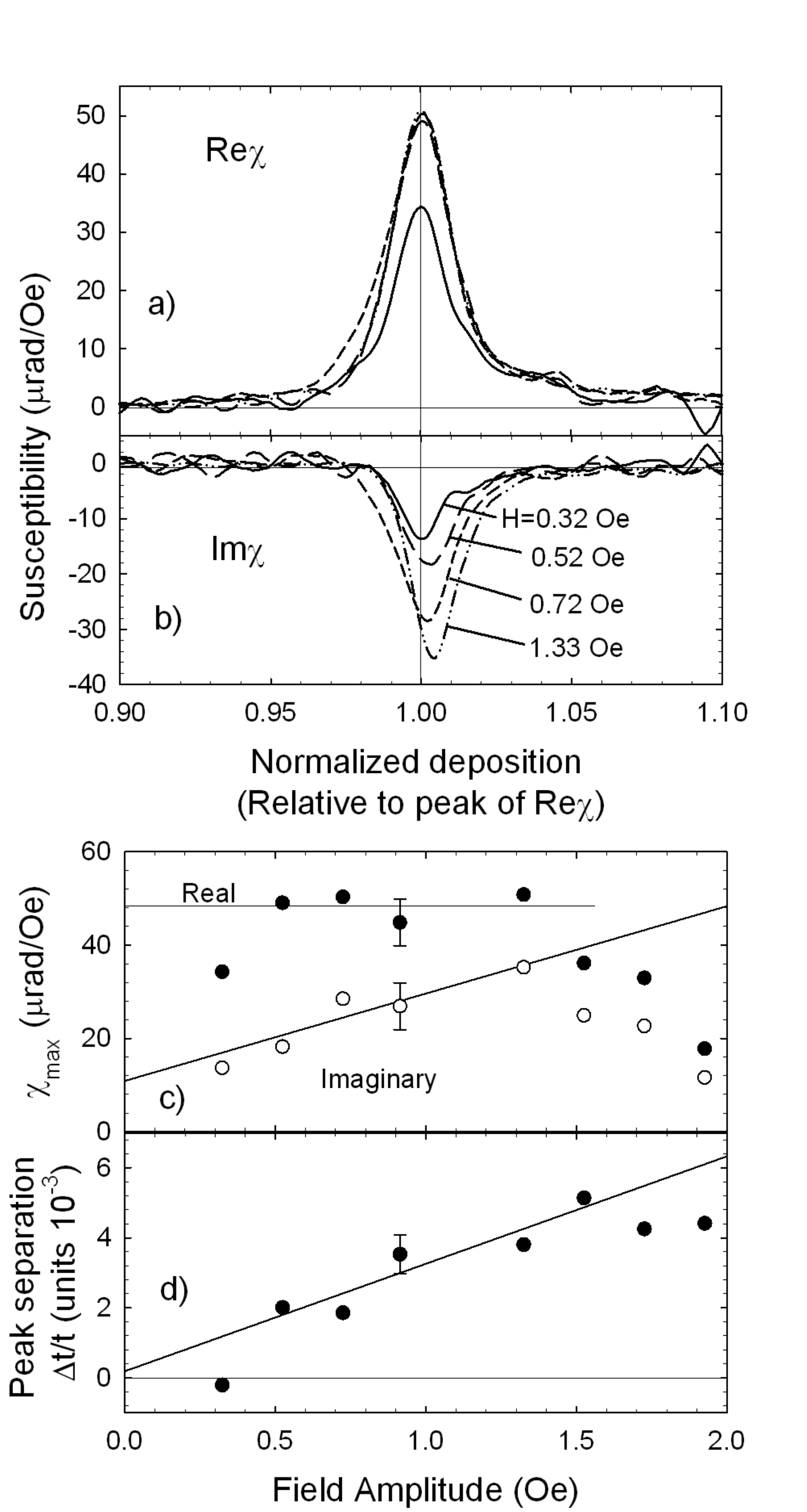}}
\caption{A series of susceptibility measurements as films were grown at 255 K, more quickly than in fig.(\ref{sample_curve}), using different field amplitudes $H$ but the same deposition rate.  The Real (a) and Imaginary (b) components of selected curves are plotted against a scale normalized to the position of the peak in Re$\chi$.  The field amplitudes are given for each curve.  c) The maximum value of the Re and Im components are plotted against the applied field, using solid and open symbols, respectively.  For $H\leq 1.33$ Oe, the linear susceptibility is measured.  The imaginary component increases linearly from a non-zero limit at $H=0$, indicating additional losses due to domain wall motion.  d) The normalized separation of the peak locations of the Re and Im components as a function of the applied field.}
\label{H_dependence}
\end{figure}

\subsection{Determining the range of scaling}
The present experiments measure the susceptibility with a finite field that drives the system at a frequency of 210 Hz.  The deposition range over which eq.(\ref{zeroT}) will apply depends upon the values of the field and measurement frequency.

The range of power law scaling of Re$\chi$ for a representative data set can be seen in the ln-ln plot in fig.(\ref{sample_curve}b), where it is marked by vertical dashed lines.  On the right-hand (lower deposition) side of the range, the signal diminishes and noise becomes noticeable.  At some point, a noise fluctuation reduces the signal to such an extent that $\ln[\mathrm{Re}\chi]$ falls dramatically and marks the end of the useful data range.  This limit could be, in principle, extended by increasing the amplitude of the applied field and, with it, the signal.  However, too large a field would not satisfy eq.(\ref{H_limit}) for the scaling range of $\rho^*_T$.

In order establish the allowable field range, a series of measurements of $\chi$ using different applied field amplitudes, $H$, were made on films as they were grown at a relatively quick rate of deposition.  Fig.(\ref{H_dependence}a) and (b) show a representative sample of these measurements, each labeled by the field amplitude. In fig.(\ref{H_dependence}c) the maximum values of Re$\chi$ and of Im$\chi$ are plotted as a function of $H$, using solid symbols and open symbols, respectively.  Each symbol represents a single film, except for the points at 0.92 Oe, where the error bar gives the standard deviation of the peak height measured for 6 films.  This error estimate is expected to apply proportionately to the measurements made with other field amplitudes.  Concentrating for the moment on Re$\chi$, it can be seen that the peak height is independent of the field amplitude for $H \leq $1.3 Oe, indicating that eq.(\ref{H_limit}) is satisfied at $\rho^*_T=\rho^*_{T,max}$, and therefore at the larger values of $\rho^*_T$ within the scaling region of fig.(\ref{sample_curve}b).  Also, comparison with eq.(\ref{delta}) indicates that the susceptibility is not limited by the use of too large a field.  The departure from linearity for field amplitudes $H \leq 0.35$ Oe is due to the loss of signal compared to the noise level.  These are the limits within which the field can be varied, and it is clear that is not possible to extend the limit of scaling by increasing the field.  As a result, the data in fig.(\ref{sample_curve}), and all the data that is analyzed later in this article, were taken in the linear region with $H=$0.92 Oe.

On the left-hand  (higher deposition) side of fig.(\ref{sample_curve}b), the limit of scaling is determined by the dynamics of the paramagnetic system.  This is illustrated in fig.(\ref{sample_curve}c), where $\omega \tau$ is presented on a ln-ln plot in accordance with eq.(\ref{zednu}) and (\ref{quotient}).  
An apparent region of dynamic scaling is indicated by the vertical solid lines, and clearly begins just as the region of scaling in Re$\chi$ ends.  The changeover occurs when $(\omega \tau)^2 \approx 0.05 - 0.07$, consistent with earlier studies of the critical susceptibility at the thermal transition\cite{Dunlavy2}.  In principle, the scaling region can be moved closer to $p^*(T)$ by reducing the measurement frequency.  In practice, the frequency of 210 Hz was chosen to avoid mechanical resonances of the apparatus and a noise floor due to 1/f noise, and cannot be reduced substantially on a logarithmic scale.  

With the range of scaling determined by field and frequency limitations of the experiment, the most objective and straight-forward method to proceed is to make a ln-ln plot of Re$\chi$, such as in fig.(\ref{sample_curve}b), and perform a multivariate least-squares fit for the three correlated parameters $\chi_+$, $\gamma_p$, and $t^*$.  Unfortunately, for the present measurements on a percolating system, least-squares fitting does not find a global minimum for all three correlated parameters.  Because the scaling region in Re$\chi$ is relatively far from $t^*$, it turns out that a statistically better fit is always found by pushing the fitted $t^*$ to higher and higher values that are clearly unphysical, with a corresponding adjustment in $\chi_+$ and $\gamma_p$.

\subsection{Determining bounds on the percolation threshold}
The failure of a purely statistical method to determine $t^*$ does not mean that nothing is known about it.  There are also constraints on its value that arise from further consideration of finite field and finite size effects.  In the following it is demonstrated that these effects influence the coverages where the maxima of Re$\chi$ and Im$\chi$ occur, and that these peak positions set bounds on $t^*$.  Then the value of $\gamma_p$ can be bounded by two separate least-squares fits for two parameters.

First consider the real part of the susceptibility.  As was discussed previously in reference to eq.(\ref{scale1}) and (\ref{H_limit}), finite size effects displace the maximum of Re${\chi}$ into the paramagnetic phase. These finite size effects are commonly encountered, and create a ``tail" in the magnetization curve into the paramagnetic phase.\cite{Elmers6}  This phenomenon is well understood from simulations of phase transitions on a finite lattice\cite{Newman}.  A careful study of the thermal transition on a high quality Fe/W(110) film by Elmers \textit{et al.}\cite{Elmers3} showed even in this case a displacement of the peak of $\chi(T)$ by $\Delta T/T_c \approx 2\times 10^{-3}$ into the paramagnetic phase. 

The use of a finite field for the susceptibility measurements also limits the divergence of $\chi_{eq}(\rho^*_T, H)$.  As is seen in eq.(\ref{rhomax}), the maximum of the equilibrium susceptibility occurs in the paramagnetic phase.  This phenomena is more familiar in measurements of the magnetization curve.   For example, a detailed study of the thermal transition in Fe/W(110) by Back \textit{et al.}\cite{Back} illustrates how a static d.c. field displaces the point of inflection in $M(T)$ into the paramagnetic phase.  For a field of about 1 Oe, the displacement is again of order $\Delta T/T_c \approx 10^{-3}$.   

Finally, as will be discussed below, there are domains in the ferromagnetic state that are not considered in the description in section IIIA. \cite{Aharoni}.  The onset of the driven dynamics of the domain walls by the finite field significantly increases the dissipation as the system moves through the transition with increasing deposition.  This new source of dissipation increases the phase angle, $\alpha$, of the lag between the Re and Im parts of the susceptibility.  This preferentially reduces Re$\chi$ in the ferromagnetic state compared to the paramagnetic state and reinforces the effects of the finite size and finite field.

These factors are difficult to quantify individually.  In particular, the use of a low frequency a.c field will produce some averaged effect compared to the static fields considered in eq.(\ref{H_limit}) and (\ref{rhomax}).  However, the main point for the present purposes is that all the factors work in the same direction to produce a maximum in Re$\chi$ within the paramagnetic phase.  The deposition at this maximum can thus be used as a lower bound on the deposition at the transition point.

Now consider the imaginary component of the susceptibility.  Non-equilibrium effects characterized by the phase lag $\alpha$ show up directly as Im$\chi$, and cause it to have a peak position that  is different than Re$\chi$.  Im$\chi$ represents dissipation, and dissipation is greater in the ferromagnetic state than in the paramagnetic state due to the driven dynamics of magnetic domains\cite{Aharoni}.  This can be seen in fig.(\ref{H_dependence}c).  In the range of applied field amplitudes where Re$\chi$ is independent of the field, Im$\chi$ grows linearly with the applied field, with a projected non-zero peak height at zero field amplitude.    The imaginary response at the limit of zero applied field is a pure phase lag because the paramagnetic system cannot follow the applied field quickly enough.  The linear increase with $H$ from this point represents the additional contribution of hysteresis losses in the ferromagnetic phase\cite{Bozorth,Rayleigh}.  For a given field amplitude, Im$\chi$ increases through the transition as the size of the minor loop that can be traversed increases.  Further into the ferromagnetic state, the growth of the coercive field overcomes the applied field, and Im$\chi$ falls, forming a peak.  As a result, the deposition where the peak of Im$\chi$ occurs is within the ferromagnetic state, and can be used as an upper bound on the deposition at the transition point.

These two bounds are consistent with previous determinations of the critical exponents of the thermal transition in Fe/W(110), where the critical temperature could be more accurately determined by statistical measures of fitting\cite{Dunlavy,Dunlavy2}.  They are confirmed for the present study in fig.(\ref{H_dependence}d).  Since a larger coercive field is required to overcome a larger applied field, the peak of Im$\chi$ is expected to move further into the ferromagnetic state as the applied field amplitude increases.  Similarly, eq.(\ref{rhomax}) shows that a larger field amplitude moves the peak of Re$\chi$ further into the paramagnetic phase.  The top $\approx$1/4 of the peaks in the data set in parts a) and b) of figure(\ref{H_dependence}) were fitted to a cubic polynomial to find the positions of the maxima.  (An example of the method is seen in the insets to fig.(\ref{sample_curve}a).)  The separation of the peak positions for Re$\chi$ and Im$\chi$ are then plotted in normalized units against the amplitude of the applied field. Because the film depositions are not calibrated absolutely, the shifts are referenced to the peak of Re$\chi$ for each curve.  Each point in fig.(\ref{H_dependence}d) represents a single experiment, except for at 0.92 Oe, where the error bar is calculated from six different films.  This error estimate is expected to apply proportionately to all the points.  It can be seen that the separation of the peak positions does in fact increase with $H$, as expected.  

With $t^*$ bounded by the locations of the two peaks, it is possible to use two two-parameter fits to calculate bounds on $\gamma_p$.

\subsection{Experimental determination of $\gamma_p$}
A series of susceptibility measurements were made on films as they were grown at 255 K.   In a previous investigation of the phase transition line for this system,\cite{Belanger} quantitative fitting of the phase transition line  showed that $\theta^*_T-\theta_c$=0.006 ML at this temperature, confirming that the measurements are made close to the limit of the $T=0$ percolation transition. The deposition rate was about 0.05 ML/min, calculated after the fact.

\begin{figure}
\scalebox{.55}{\includegraphics{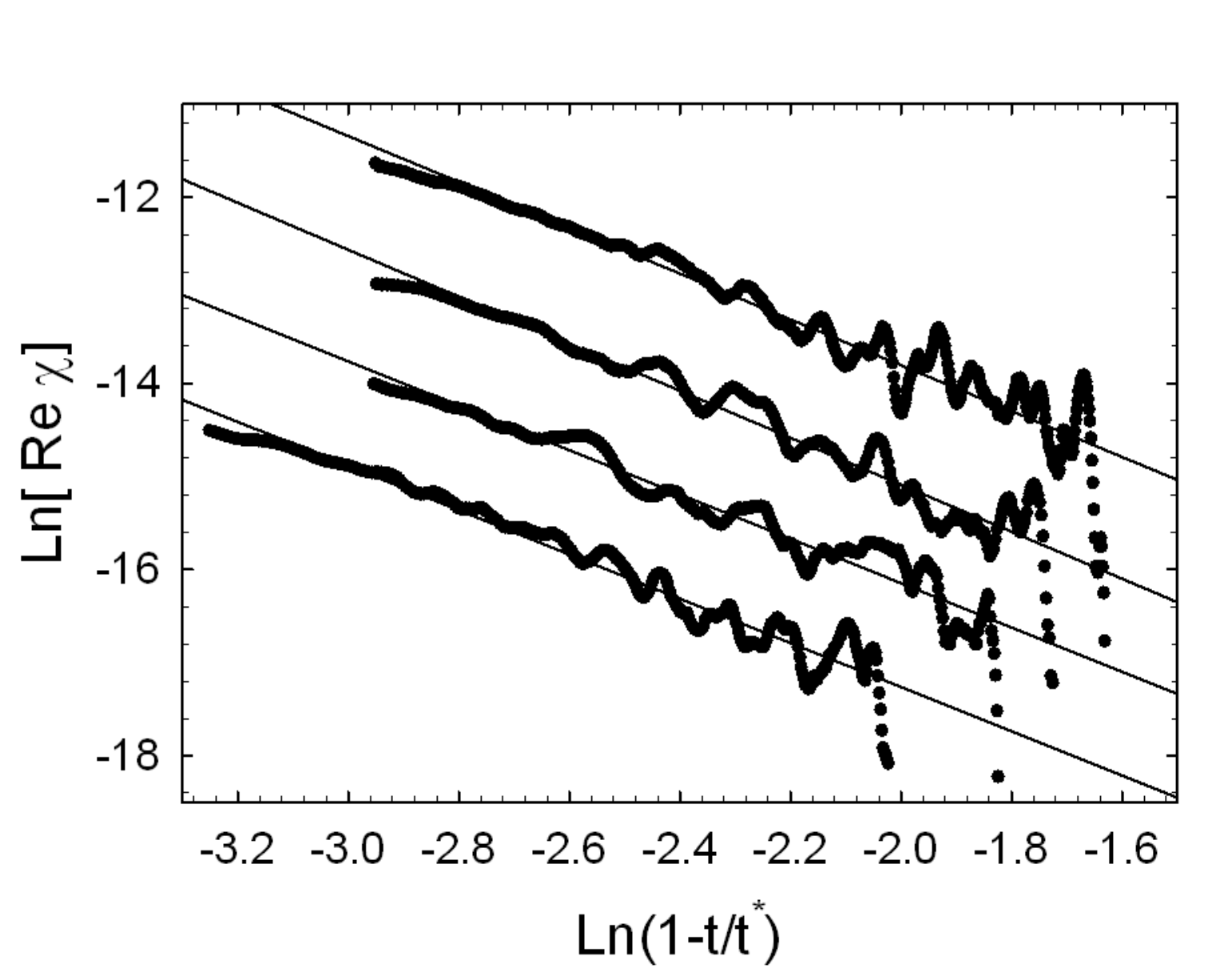}}
\caption{Four data sets (one of which is shown in fig.(\ref{sample_curve})), measured at a slow growth rate at 255 K, are plotted on a ln-ln scale.  The curves are displaced vertically for clarity.  The left end of the fitted regions are determined by $(\omega \tau)^2=0.06$, and the right end by the obvious influence of noise that causes Re$\chi$ to become negative (the resulting trailing tails seen on the plot are not included in the fits).   Parameters from the fitted lines are presented in Table \ref{fitted}.}
\label{ln_fits}
\end{figure}

The four data sets that were suitable for further analysis are presented in fig.(\ref{ln_fits}) in a logarithmic plot.  The main difficulty was obtaining data sets with a range of at least 1.0 in $\ln(1-t/t^*)$.  The curves in fig.(\ref{ln_fits}) extend on the left-hand side to the limit of  $(\omega\tau)^2=0.06$ set by the dynamics of the paramagnet.  Obtaining a satisfactory range in $\ln(1-t/t^*)$ then depended on extending the right hand limit as far as possible into the paramagnetic tail of Re$\chi$.  As can be seen in fig.(\ref{sample_curve}a), this in turn depended upon the random nature of the measurement noise, so that on occasion a large noise fluctuation did \textit{not} occur during this crucial part of the measurement, and the data set was acceptable.  The traces in fig.(\ref{ln_fits}) show for illustration that eventually a noise fluctuation occurred and ended the scaling region.  These large noise excursions are not included in the subsequent fitting for $\gamma_p$.

\begin{table}
\caption{\label{fitted}The parameters for the linear fits in fig.(\ref{ln_fits}).  There are two least-squares fits to eq.(\ref{zeroT}) on logarithmic scales to define bounds on the parameters.  One fit uses $t^*=t^*_+$ at the peak of Re$\chi(t)$; the other fit uses $t^*=t^*_-$ at the peak of Im$\chi(t)$.}
\begin{ruledtabular}
\begin{tabular}{|l|l|l|l|l|}
Data set          &         1     &     2      &     3    &   4     \\ \hline
$\gamma_p \pm0.03$ @ $t^*_+$     & 2.340     &  2.318 &  2.252 &  2.313  \\ \hline
$\gamma_p \pm0.03$ @ $t^*_-$     &  2.516    &  2.470 & 2.389 & 2.536  \\ \hline
mean   & 2.428   & 2.394   & 2.321   & 2.425 \\ \hline
bounded range   & $\pm0.088$  &  $\pm0.076$  &   $\pm0.069$ & $\pm0.112 $ \\ \hline
average &\multicolumn{4}{l|}{$\gamma_p = 2.39\pm0.04$} \\ \hline \hline
$\ln(\chi_+) \pm0.1$ @ $t^*_+$    & -18.90     &   -18.75  & -18.29 & -18.84  \\ \hline
$\ln(\chi_+) \pm0.1$ @ $t^*_-$    & -19.10    &  -18.97   &  -18.48 &  -19.20 \\  \hline
mean & -19.00 &  -18.86  &  -18.39  & -19.02 \\ \hline
bounded range & $\pm0.10$ & $\pm0.11$ & $\pm0.10$ & $\pm0.18$ \\ \hline
average & \multicolumn{4}{l|}{$\ln(\chi_+)$=-18.82$\pm0.06$}
% Lines of table here ending with \\
\end{tabular}
\end{ruledtabular}
\end{table}

The value of $t^*$ used to plot fig.(\ref{ln_fits}) is midway between the deposition where the peaks in Re$\chi$ and Im$\chi$ occur.  The results for least squares fitting for $\gamma_p$ and the amplitude $\chi_+$ are presented in Table I.  Values are given for fits using the values $t^*=t^*_+$ at the peak of Re$\chi$ and $t^*=t^*_-$ at the peak of Im$\chi$.  The stated uncertainty $\pm0.03$ in the first column is due the individual, two parameter least-squares fitting alone. Subsequent rows in the Table give the mean value of $\gamma_p$ and the range of $\gamma_p$ between these bounds for each data set.  Since the bounded ranges of $\gamma_p$ for all four data sets overlap, and are similar in size, a final average value is obtained from simple averaging of the four means and reducing the range by $\sqrt{N}=2.$  This yields $\gamma_p=2.39\pm0.04$.  This result is in agreement with the theoretical value\cite{StaufferAharony} of 43/18=2.388...

\section{Conclusions}
The magnetic phase transition that occurs as the deposition of second layer of Fe atoms on W(110) reaches a critical value provides a unusual opportunity to study the 2D percolation transition.  Isolated second layer islands that are perpendicularly magnetized are coupled antiferromagnetically \textit{via} the first layer of Fe and the W substrate, by mechanisms that are not well understood.  This produces a magnetically frustrated state with no long-range order.  During percolation of the second layer, long-range in-plane order arises and can be detected as a narrow, robust peak in the magnetic susceptibility, as is expected for a second order transition.  Although questions remain at the microscopic level, experimental studies have shown that the transition is described quantitatively by the universal characteristics of a 2D percolation transition at finite temperature.  This includes the quantitative description of the transition line in the $(p,T)$ plane, and the quantitative form of the paramagnetic susceptibility near percolation, $\chi(\rho^*_T \approx 0, T)$, measured as a function of temperature\cite{Belanger}.  The present article adds to this characterization the determination of the critical exponent of the magnetic susceptibility (mean island size) as $\gamma_p=2.39\pm0.04$, in agreement with the theory of 2D percolation.  This measurement, in conjunction with the previous measurements of $\chi(\rho^*_T \approx 0, T)$, are consistent with a crossover exponent of $\phi=1$.  An definitive experimental finding for $\phi$ is not possible because of the experimental uncertainty in the absolute deposition for the measurements as a function of temperature.\cite{Belanger}

The particular difficulties that arise in the experimental determination of static critical exponents were addressed using objective criteria based upon the fact that the experiments were performed using a finite field amplitude oscillating at a finite frequency, and that the percolated sample is not uniform.  These effects made it possible to determine bounds on the percolation deposition, rather than fitting for it.  All relevant parameters are determined independently from within each data set with no uncertainty due to absolute calibration, and the four acceptable data sets show reproducible results.  These procedures have allowed the evaluation of a small  uncertainty of 1.7\% on the experimentally determined value of $\gamma_p$.  We are unaware of a previous experimental measurement of a static 2D percolation critical exponent as a function of deposition/dilution that meets similar criteria.   

Although the universal static behaviour of this percolation transition is established, there remain interesting questions at the microscopic level.  On the one hand, the transition conforms to the description of percolation of a 2D Ising system at finite temperature, where the properties of fragile 1D Ising chains linking larger islands play a crucial role in the critical region for percolation\cite{Coniglio}.  On the other hand, the frustrated magnetic state from which the long-range in-plane ferromagnetic state derives upon percolation appears to be mediated by the continuous basal layer of Fe atoms and W substrate even before percolation occurs\cite{Elmers2}.  While it is not yet clear whether or not this is a contradiction, it is certainly significant to the dynamics of the system.  This can be seen from the fact that the transition due to the percolation of the \textit{second} layer Fe atoms is detected using low-field susceptibility measurements as the films are grown (even in experiments below 230 K), but the more conventional transition due to the percolation of \textit{first} layer Fe atoms is not.  Insight into these questions might be found by investigating the critical slowing down, which is dependent on the underlying dynamics of the system.

%\appendix
%\section{}

\begin{acknowledgments}
The authors would like to thank M. Kiela for continuing technical contributions, and E. Sorensen for helpful discussions.  Support from the Natural Sciences and Engineering Research Council of Canada is gratefully acknowledged.
\end{acknowledgments}

\bibliography{percolation}

\end{document}